\documentstyle[twocolumn,prl,aps]{revtex}
\begin{document}
\draft

\def \d {{\rm d}}

\font\grb=eurb10
\font\ccc=cmr12
\def\bphi{\hbox{\grb\char'047}\,}
\def\bpsi{\hbox{\grb\char'040}\,}
\def\bchi{\hbox{\grb\char'037}\,}
\def\bfi{\hbox{\grb\char'036}\,}
\def\varkap{\hbox{\ccc\char'032}\,}

\def\A{\hbox{\bf A}}
\def\B{\hbox{\bf B}}
\def\a{\hbox{\bf a}}
\def\b{\hbox{\bf b}}
\def\c{\hbox{\bf c}}
\def\e{\hbox{\bf e}}
\def\f{\hbox{\bf f}}
\def\k{\hbox{\bf k}}
\def\l{\hbox{\bf l}}
\def\m{\hbox{\bf m}}
\def\n{\hbox{\bf n}}
\def\p{\hbox{\bf p}}
\def\q{\hbox{\bf q}}
\def\u{\hbox{\bf u}}
\def\v{\hbox{\bf v}}
\def\M{\hbox{\bf M}}
\def\N{\hbox{\bf N}}
\def\R{\hbox{\bf R}}
\def\U{\hbox{\bf U}}
\def\V{\hbox{\bf V}}
\def\W{\hbox{\bf W}}
\def\bigpsi{{\bf \Psi}}
\def\bigomega{{\bf \Omega}}

\title{Solving the characteristic initial value problem for
colliding plane\\ gravitational and electromagnetic waves}

\author{G. A. Alekseev$^1$
  \thanks{E-mail: {\tt G.A.Alekseev@mi.ras.ru}} \
and J. B. Griffiths$^2$
  \thanks{E--mail: {\tt J.B.Griffiths@Lboro.ac.uk}}}
\address{$^1$Steklov Mathematical Institute,
Gubkina 8, Moscow 117966, GSP-1, Moscow, Russia.
\\
$^2$Department of Mathematical Sciences, Loughborough
University, Loughborough, Leics. LE11 3TU, U.K. \\ }

\date{\today}

\maketitle

\begin{abstract}
\noindent A method is presented for solving the characteristic
initial value problem for the collision and subsequent nonlinear
interaction of plane gravitational or gravitational and
electromagnetic waves in a Minkowski background. This method
generalizes the monodromy transform approach to fields with
nonanalytic behaviour on the characteristics inherent to waves
with distinct wave fronts. The crux of the method is in a
reformulation of the main nonlinear symmetry reduced
field equations as linear integral equations whose solutions
are determined by generalized (``dynamical'') monodromy data
which evolve from data specified on the initial characteristics
(the wavefronts).
\end{abstract}

\pacs{04.40.-b, 04.30.-w}

\narrowtext

\section{Introduction}

The collision and subsequent nonlinear interaction between plane
gravitational or gravitational and electromagnetic waves propagating
with distinct wavefronts in a Minkowski background is a well
formulated characteristic initial value problem. However, even the
discovery of the integrability of the main field equations for this
situation did not lead to a solution of this complex nonlinear
problem. None of the existing solution generating methods have been
found suitable for an effective construction of the solution in the
wave interaction region starting from given characteristic initial
data determined by the approaching waves.

The structure of the governing field equations for colliding plane
waves, physical and geometrical interpretations, and various
particular solutions and techniques, have been described in
\cite{Griffiths:1991}. For colliding plane gravitational waves with
aligned constant polarizations, the vacuum Einstein equations are
reducible to the linear Euler-Poisson-Darboux equation. In this
case, the corresponding characteristic initial value problem can be
solved using the generalized version of Abel's transform
\cite{Hauser-Ernst:1989}.

However, when the polarizations of the approaching gravitational
waves are not constant and aligned, or in the presence of
electromagnetic waves, the governing equations are essentially
nonlinear and are equivalent to the hyperbolic form of the Ernst
equations. For this case, an appreciable number of particular
solutions are known. These have been found using the ``inverse''
method in which a formal solution in the interaction region is first
constructed and the corresponding characteristic initial data for
the approaching waves is only determined subsequently. Recently,
infinite hierarchies of exact vacuum and electrovacuum solutions
with an arbitrary number of free parameters were found
\cite{Alekseev-Griffiths:2000}, and many of these are of the type
appropriate for colliding plane waves. However, it is not a simple
technical task to simplify these solutions for particular cases and
to calculate the corresponding characteristic initial data.

For the analysis of the characteristic initial value problem for the
vacuum hyperbolic Ernst equation, Hauser and Ernst
\cite{Hauser-Ernst:1990-1991} have generalized their
group-theoretical approach (which had been developed earlier for
stationary axisymmetric fields) and constructed a homogeneous
Hilbert problem with corresponding matrix linear integral equations.
Many aspects of this problem, including the existence and uniqueness
of solutions and a detailed proof of the Geroch conjecture, were
elaborated in \cite{Hauser-Ernst:2001}.

However, a general scheme for the solution of various nonlinear
initial and boundary value problems for integrable reductions of
Einstein's equations had been developed in the framework of the
monodromy transform approach
\cite{Alekseev:1985}--\cite{Alekseev:2000}. This can be applied
to both the characteristic initial value problem and the Cauchy
problem for the hyperbolic case, as well as some boundary
problems for the elliptic case. In this approach, every solution can
be characterized by a set of functions of an auxiliary (spectral)
parameter. These functions are interpreted as the monodromy data on
the spectral plane of the fundamental solution of an auxiliary
overdetermined linear system associated with the (nonlinear) field
equations. These monodromy data are nonevolving (i.e.  coordinate
independent) and, generally, can be chosen arbitrarily or
specified in accordance with the properties of the solution being
sought. In particular, these data can be determined (at least in
principle) from the initial or boundary data. In this scheme, the
solution of the initial or boundary value problem is determined by
the solution of some linear singular integral equations whose kernel
is constructed using these (specified) monodromy data.

Recently a further generalization of this approach was derived in
\cite{Alekseev:2001}, in which a new linear integral equation form
of various hyperbolic integrable reductions of Einstein equations
was constructed.  The scalar kernel of these quasi-Fredholm
equations depends on the monodromy data in a different way --
referred to as ``dynamical monodromy data''. Unlike the previous
method, these data evolve, and their evolution is prescribed by the
characteristic initial conditions.  It appears that these new
integral equations are better adapted to the construction of an
effective solution of initial value problems because their
coefficients carry more explicit information about the characteristic
initial data and the corresponding analytical structures of the
solutions desired.

In this paper, we present a new approach to the solution of the
colliding plane wave problem and also describe a method that can be
implemented in practice to derive explicit solutions. Our
construction generalizes the monodromy transform approach whose
formulation in the above mentioned papers was applicable only for
fields that are analytically dependent on some special set of
geometrically defined space-time coordinates. (One of these
coordinates determines the measure of an area on the two-dimensional
orbits of the space-time isometry group. The other is its harmonic
conjugate.) However, for colliding plane waves propagating with
distinct wave fronts on a Minkowski background, this analyticity is
obviously violated on the wave fronts. Moreover, the physically
accepted matching conditions (the ``colliding plane wave
conditions'') imply regular behaviour of the field components
near the wave fronts in some appropriate null coordinates
\cite{Griffiths:1991}. In terms of geometrically defined
coordinates, this regularity leads to a specific
$\underline{\mbox{\it singular}}$ behaviour of the coefficients of
the associated linear system on these hypersurfaces, where the first
derivatives of the field components become infinite. This gives rise
to crucial consequences for the previous formulation.  Additional
singularities appear on the spectral plane for some auxiliary
functions. Also, the integral representations diverge at the
singular points, while the normalization of the solutions remains a
necessity. Another important phenomenon which arises in these
singular cases is that the nonevolving monodromy data, which
continue to exist, lose their the most important property -- their
unambiguous characterization of the solution.

The solution of these problems arises from our recent observation
that one of two linear integral ``evolution equation'' forms of the
field equations derived in \cite{Alekseev:2001} admits a
generalization to the singular case. It is also important that the
dynamical monodromy data can still be used in this case to
characterize the solutions. In this paper we present the generalized
linear quasi-Fredholm integral ``evolution equation'', which covers
the singular case and opens a direct way for the construction of
solutions for colliding plane waves from given characteristic
initial data. We also briefly discuss some applications.

\section{Associated linear system with spectral parameter}

We base our construction on the Kinnersley-like overdetermined
linear system with a free complex parameter $w$ for a $N\times
N$ matrix function $\bigpsi(\xi,\eta,w)$ whose integrability
conditions are equivalent to the hyperbolic space-time symmetry
reduction for $N=2$ of the vacuum Einstein equations and for
$N=3$ of the electrovacuum Einstein--Maxwell equations --
  $$ \left\{\begin{array}{lclcl}
  2i(w-\xi)\partial_\xi\bigpsi= \U(\xi,\eta)\bigpsi,&&
  \mbox{rank}\,\U=1, && \mbox{tr}\,\U=i,\\
  2i(w-\eta)\partial_\eta\bigpsi= \V(\xi,\eta)\bigpsi,&&
  \mbox{rank}\,\V=1, && \mbox{tr}\,\V=i
   \end{array} \right. $$
in which the null coordinates  $\xi$, $\eta$ are certain
linear combinations of geometrically defined nonnull
coordinates mentioned in the introduction. These
equations must be supplemented with the additional
constraints on its matrix integral
\cite{Alekseev:1987,Alekseev:2000}
  $$ \left\{ \begin{array}{l}
\bigpsi^\dagger\>\W\> \bigpsi = \W_0(w),\\
\W_0^\dagger(w)=\W_0(w),
   \end{array}
   \quad {\partial\W\over \partial w}=4 i\bigomega
\right. $$
  where ${}^\dagger$ denotes Hermitian conjugation
($\W_0^\dagger(w)\equiv\overline{\W^T_0(\overline{w})}$) and
$\bigomega$ is a constant matrix. For $N=2,3$, the only nonzero
components of $\bigomega$ are $\bigomega^{12}=1$ and
$\bigomega^{21}=-1$. For $N=3$, the condition $\W^{33}=1$ should
be also satisfied.  The real null coordinates  $\xi$,
$\eta$ are certain linear combinations of geometrically defined
nonnull coordinates, say $\alpha$, $\beta$, mentioned in
the introduction.

For any solution of these conditions, the components of $\U$, $\V$
and $\W$ can be identified with certain metric components and the
electromagnetic potential and their derivatives. Without loss of
generality, we impose the normalization conditions at the point
denoted by $(\xi_\times,\eta_\times)$ at which the waves collide.
These are $\bigpsi(\xi_\times,\eta_\times,w) = \mbox{\bf I}$ and
$\W_0(w)=4 i(w-\beta_\times)\bigomega
+\mbox{diag}\,(4\alpha_\times^2,4,1)$, where
$\alpha_\times=(\xi_\times-\eta_\times)/2$ and
$\beta_\times=(\xi_\times+\eta_\times)/2$. A shift of origin and a
rescaling of the coordinates $\xi$ and $\eta$ allow us to specify
below $\xi_\times=1$ and $\eta_\times=-1$, so that
$\alpha_\times=1$, $\beta_\times=0$.

\section{The colliding plane wave problem}

For plane gravitational or gravitational and electromagnetic waves
with distinct wavefronts which collide in a Minkowski background, it
is well known \cite{Griffiths:1991} that there exist global null
coordinates $(u,v)$ such that the O'Brien-Singe matching conditions
imposed on the wavefronts $u=0$ and $v=0$ lead to a well posed
characteristic initial value problem. However, the regularity of
this problem in these global coordinates in the interaction region
($u\ge 0$, $v\ge 0$) implies special relations between these
coordinates and the coordinates ($\xi$, $\eta$) which are
nonanalytical on the boundaries.  It is possible to use the
coordinate freedom to put $$ \xi=1-2 u^{n_+}, \qquad \eta=-1+2
v^{n_-}, $$ where $n_\pm \ge 2$ are specified as part of the initial
data. Thus, the important specific of this characteristic initial
value problem is that, in terms of geometrically defined coordinates
$(\xi,\eta)$, the first derivatives of the field components should
be discontinuous and even unbounded on the wavefronts and at the
point of collision.

\section{The analytical structure of $\bigpsi$ on the spectral
plane}

Everywhere below, $\bigpsi$ denotes the fundamental solution of the
associated linear system described above which is normalized at the
point of collision ($u=0,v=0$, i.e. ($\xi=1,\eta=-1$). As in the
regular case \cite{Alekseev:1987,Alekseev:2000}, the structure of
the coefficients of the associated linear system for $\bigpsi$
implies that the normalized fundamental solution
$\bigpsi(\xi,\eta,w)$ and its inverse possess four branch points on
the spectral plane $w$, namely at $w=\xi$, $w=\eta$, $w=1$ and
$w=-1$. The order of these points and our choice of the cuts $L_\pm$
joining them are indicated in the figure $$\matrix{
L_{\scriptscriptstyle-} \hskip16ex L_{\scriptscriptstyle+} \cr
\noalign{\vskip-1.5ex} {\vrule width10ex height0.05ex depth0.05ex}
{\vrule width0.05ex height0.5ex depth0.5ex} {\vrule width10ex
height0.15ex depth0.15ex} {\vrule width0.05ex height0.5ex
depth0.5ex} {\vrule width10ex height0.05ex depth0.05ex} {\vrule
width0.05ex height0.5ex depth0.5ex} {\vrule width10ex height0.15ex
depth0.15ex} {\vrule width0.05ex height0.5ex depth0.5ex} {\vrule
width10ex height0.05ex depth0.05ex}\cr \noalign{\vskip-0.5ex} -1
\hskip8ex \eta \hskip10ex \xi \hskip8ex 1}$$ Near these singular
points, and on the cuts $L_\pm$, the components of $\bigpsi$ and its
inverse (as functions of $w$ for given $u\ge0$ and v$\ge 0$) possess
in general the local structure: $$\bigpsi(\xi,\eta,w)=
\widetilde{\bpsi}_\pm(\xi,\eta,w) \otimes\k_\pm(w)
+\M_\pm(\xi,\eta,w) $$ where, as in the regular case, the coordinate
independent components of the row vectors $\k_\pm(w)$ constitute the
``projective vectors'' of the monodromy data. Their components, as
well as the components of the matrices $\M_\pm(\xi,\eta,w)$, are
regular (holomorphic) on the cuts $L_\pm$ with the corresponding
index. In the analytical case  we always have
$\widetilde{\bpsi}_+(\xi,\eta,w) =\sqrt{w-1\over w-\xi}
\bpsi_+(\xi,\eta,w)$ and
$\widetilde{\bpsi}_-(\xi,\eta,w)=\sqrt{w+1\over w-\eta}
\bpsi_-(\xi,\eta,w)$ where the vectors $\bpsi_+$ and $\bpsi_-$ are
holomorphic on the cuts $L_+$ and $L_-$ respectively. In the general
case, the components of the column vectors
$\widetilde{\bpsi}_\pm(\xi,\eta,w)$ also have branch points at the
endpoints of the corresponding cuts $L_\pm$, but the character of
their singularities at the points $w=1$ and $w=-1$ respectively (as
well as the powers $n_+$ and $n_-$ defined above) are determined by
the initial data.

\section{The integral ``evolution'' equations and the solution of
the problem}

To set up the characteristic initial value problem, we introduce two
matrix functions which are the normalized fundamental solutions of
the linear ordinary differential equations -- the restrictions of
the associated linear system to the characteristics $\xi=1$ and
$\eta=-1$: $$\begin{array}{ll} 2i(w-\xi)\partial_\xi\bigpsi_+
=\U(\xi,-1)\cdot\bigpsi_+, \qquad &
\bigpsi_+(1,w)=\hbox{\bf I}\\
2i(w-\eta)\partial_\eta\bigpsi_-=\V(1,\eta) \cdot\bigpsi_-, \qquad &
\bigpsi_-(-1,w)=\hbox{\bf I}
\end{array}$$
in which the coefficients are determined by the initial data for the
fields on the corresponding characteristics.  These matrices should
be the characteristic initial data for the required solution for
$\bigpsi(\xi,\eta,w)$: $$\bigpsi_+(\xi,w)\equiv\bigpsi(\xi,-1,w),
\quad \bigpsi_-(\eta,w)\equiv\bigpsi(1,\eta,w) \label{idata} $$

The analytical structures of $\bigpsi_\pm$ on the spectral plane
are very similar to those of $\bigpsi$. Namely,
$\bigpsi_\pm(w=\infty)={\bf I}$, $\bigpsi_+$ is holomorphic
outside $L_+$ and $\bigpsi_-$ outside $L_-$, and their local
  structures on these cuts are given by $$\begin{array}{ll}
   L_+:\quad\bigpsi_+(\xi,w)=
\widetilde{\bpsi}_{0+}(\xi,w)\otimes\k_+(w)+\M_{0+}(\xi,w),\\
  L_-:\quad\bigpsi_-(\eta,w)=
\widetilde{\bpsi}_{0-}(\eta,w)\otimes\k_-(w)+\M_{0-}(\eta,w)
   \end{array}$$ where $k_\pm(w)$ are the same as for $\bigpsi$,
and $\M_{0+}(\xi,w)$ and $\M_{0-}(\eta,w)$ are holomorphic on
$L_+$ and $L_-$ respectively.

We now introduce, in analogy with the regular case
\cite{Alekseev:2001}, the ``evolution'' or ``scattering''
matrices $\bchi_\pm(\xi,\eta,w)$, representing
$\bigpsi(\xi,\eta,w)$ in two alternative forms
   \begin{eqnarray}
  &&\bigpsi(\xi,\eta,w)=\bchi_+(\xi,\eta,w)\cdot
\bigpsi_+(\xi,w) \nonumber\\
  &&\bigpsi(\xi,\eta,w)=\bchi_-(\xi,\eta,w)\cdot
\bigpsi_-(\eta,w).\nonumber
  \end{eqnarray}
The crucial point is that the components of the matrix
$\bchi_+(\xi,\eta,w)$ are holomorphic on $L_+$ and possess a
jump on $L_-$ only, while the components of the matrix
$\bchi_-(\xi,\eta,w)$ are holomorphic on $L_-$ and possess a
jump on $L_+$. From the above it is clear also, that these jumps
are represented by highly degenerate matrices and that
$\bchi_\pm(\xi,\eta,w=\infty)={\bf I}$. These properties permit
us to represent $\bchi_\pm$ as Cauchy integrals of the form
$$\begin{array}{l}\bchi_+(\xi,\eta,w)={\bf
I}+\displaystyle{{1\over \pi
i}\int\limits_{L_-}{[\widetilde{\bpsi}_-]_{\zeta_-} \otimes
\m_-(\xi,\zeta_-)\over w-\zeta_-}\, d\zeta_-}\\
\bchi_-(\xi,\eta,w)={\bf I}+\displaystyle{{1\over \pi
i}\int\limits_{L_+}{[\widetilde{\bpsi}_+]_{\zeta_+} \otimes
\m_+(\xi,\zeta_+)\over w-\zeta_+}\, d\zeta_+}\end{array}$$ where
$[\ldots]$ is the jump (a half of the difference between the left
and right limits) of a function on a cut and $$ \matrix{
\m_-(\xi,w)=\k_-(w)\cdot\bigpsi_+^{-1}(\xi,w) \cr \noalign{\medskip}
\m_+(\eta,w)=\k_+(w)\cdot\bigpsi_-^{-1}(\eta,w)} $$ represent a new
(evolving) kind of monodromy data introduced for the regular case in
\cite{Alekseev:2001} and called there the ``dynamical'' monodromy
data. It is necessary to note here our conjecture of the convergency
of the Cauchy integrals for $\bchi_\pm$ at $w=-1$ and $w=1$
respectively which is confirmed for particular examples for any
$n_\pm\ge 2$.

The above alternative representations for $\bigpsi$ should satisfy
an obvious condition $\bchi_+\bigpsi_+\equiv \bchi_-\bigpsi_-$. This
condition considered on the cuts $L_\pm$ together with the
constructed integral representations for $\bchi_\pm$ leads to the
linear integral equations (for brevity we omit here the parametric
dependence of all objects upon $\xi$ and $\eta$): $$\begin{array}{l}
\bfi_+(\tau_+)- \displaystyle{\int\limits_{L_-}} S_+(\tau_+,\zeta_-)
\bfi_-(\zeta_-) \,d\zeta_- = \bfi_{0+}(\tau_+)\\
  \bfi_-(\tau_-)-\displaystyle{\int\limits_{L_+}}
S_-(\tau_-,\zeta_+) \bfi_+(\zeta_+) \,d\zeta_+ =
  \bfi_{0-}(\tau_-) \end{array}$$ where $\tau_+,\zeta_+\in L_+$,
$\tau_-,\zeta_-\in L_-$ and the vector functions
$\bfi_+(\tau_+)$, $\bfi_-(\tau_-)$ and their initial values
$\bfi_{0+}(\tau_+)$, $\bfi_{0-}(\tau_-)$ are the jumps
$[\widetilde{\bpsi}_+]_{\tau_+}$,
$[\widetilde{\bpsi}_-]_{\tau_-}$ and
$[\widetilde{\bpsi}_{0+}]_{\tau_+}$,
$[\widetilde{\bpsi}_{0-}]_{\tau_-}$ respectively. The scalar
kernels are given by $$\begin{array}{l} S_+(\xi,\tau_+,\zeta_-)
=\displaystyle{1\over i\pi(\zeta_--\tau_+)}
\Big(\m_-(\xi,\zeta_-)\cdot \bfi_{0+}(\xi,\tau_+)\Big) \\
S_-(\eta,\tau_-,\zeta_+) =\displaystyle{1\over
i\pi(\zeta_+-\tau_-)} \Big(\m_+(\eta,\zeta_+)\cdot
\bfi_{0-}(\eta,\tau_-)\Big)\end{array}$$

As in the regular case, the coefficients of the above integral equations are determined by the initial data. However, these
generalized ``evolution equations'' possess a more complicated singular structure. They also can be easily decoupled into two independent equations for $\bfi_+$ and $\bfi_-$, but we omit these
calculations here.

Our construction of solutions for colliding plane waves begins with the constants $n_\pm\ge 2$ which determine the degree of
non-smoothness of the fields on the wavefronts
and the characteristic initial data for the fields in terms of the Ernst potentials ${\cal E}(u,v)$, $\Phi(u,v)$ which characterise every solution. These data, viz. ${\cal E}_+(u)$, $\Phi_+(u)$, ${\cal E}_-(v)$, and $\Phi_-(v)$, should be chosen to satisfy the normalization conditions ${\cal E}_\pm(0)=-1$, $\Phi_\pm(0)=0$ and two wavefront regularity conditions \cite{Griffiths:1991}: $\vert{\cal E}^\prime_\pm(0)\vert^2 +4\vert\Phi^\prime_\pm(0)\vert^2 = 8(1-{1/n_\pm})$.
With this data we have to solve the ordinary differential equations for $\bigpsi_\pm$ and the integral equations for $\bfi_\pm$. The Ernst potentials of the sought-for solution can be evaluated then as
$$\begin{array}{l} {\cal E}(u,v)={\cal E}_+(u)
+{2\over\pi}\displaystyle{ \int\limits_{L_-}}
(\e_1\cdot\bfi_-(\zeta_-))(\m_-(\zeta_-)\cdot\e_2)
d\zeta_- \\ \phantom{{\cal E}(u,v)} ={\cal
E}_-(v)+{2\over\pi}\displaystyle{ \int\limits_{L_+}}
(\e_1\cdot\bfi_+(\zeta_+))(\m_+(\zeta_+)\cdot\e_2)
d\zeta_+ \\ \Phi(u,v)=\Phi_+(u)
-{2\over\pi}\displaystyle{ \int\limits_{L_-}}
(\e_1\cdot\bfi_-(\zeta_-))(\m_-(\zeta_-)\cdot\e_3)
d\zeta_- \\
\phantom{\Phi(u,v)}=\Phi_-(v)-{2\over
\pi}\displaystyle{ \int\limits_{L_+}}
(\e_1\cdot\bfi_+(\zeta_+))(\m_+(\zeta_+)\cdot\e_3)
d\zeta_+
\end{array}$$
where $\e_1=\{1,0,0\}$, $\e_2=\{0,1,0\}$ and $\e_3=\{0,0,1\}$.

As a simple test, we consider $n_+=n_-=2$ and chose  $${\cal E}_+=-(1-u)^2,\qquad {\cal E}_-=-(1-v)^2$$
For these vacuum data our calculations lead immediately to the Khan--Penrose solution for the collision of impulsive
gravitational waves with colinear polarizations. Similar
calculations for the initial data ($\delta_\pm$ are real constants)
$${\cal E}_+=-1+2
e^{i\delta_+} u-u^2,\qquad{\cal E}_-=-1+2 e^{i\delta_-} v-v^2
$$
give the Nutku--Halil solution for the collision of these waves with noncolinear polarizations.
The initial data
$$ {\cal E}_+=-1,\quad \Phi_+=u,\qquad{\cal E}_-=-1,\quad \Phi_-=v $$
lead to the Bell--Szekeres solution for the collision of electromagnetic step waves with aligned polarizations. When the polarizations of these waves is nonaligned, we change the data above to $\Phi_-=v e^{i\gamma}$, ($\gamma$ is a real constant). This produces  $$ {\cal E}(u,v)=-1,\quad \Phi(u,v)=u\sqrt{1-v^2}+v\sqrt{1-u^2} e^{i\gamma} $$  which leads to a nontrivial (nondiagonal) metric (see \cite{Griffiths:1985}). Clearly, various new solutions can be found in the same way for different choices of the initial data.

\section*{Acknowledgments}

The work of GA was partly supported by the EPSRC, by the INTAS
grant 99-1782 and by the grants 99-01-01150, 99-02-18415 from
the RFBR.

\end{document}